\documentclass{elsart5}

 \usepackage{graphicx}

\usepackage{amssymb}

\begin{document}

\begin{frontmatter}

\title{Magnetic Excitations of the 2-D Sm Spin Layers in Sm(La,Sr)CuO$_4$}

\author[aff1]{F. Ronning\corauthref{cor1}}
\ead{fronning@lanl.gov} \corauth[cor1]{}
\author[aff1]{C. Capan}
\author[aff2]{N.O. Moreno}
\author[aff1]{J.D. Thompson}
\author[aff1]{L.N. Bulaevskii}
\author[aff1]{R. Movshovich}
\author[aff3]{D. van der Marel}
\address[aff1]{Los Alamos National Lab, Los Alamos, NM, 87545, USA}
\address[aff2]{Departamento de F\'{\i}sica, Universidade Federal de Sergipe, S\~{a}o Cristov\~{a}o - SE CEP 49100-000, Brazil}
\address[aff3]{DPMC-Geneva,
24, quai Ernest Ansermet, 1211 Geneva 4, Switzerland}
\received{12
June 2005} \revised{13 June 2005} \accepted{14 June 2005}
%use optional labels to link authors explicitly to addresses:

%\author{}
%\address{}

\begin{abstract}
 We present specific heat and susceptibility data on Sm(La,Sr)CuO$_4$ in magnetic fields up to 9~T
 and temperatures down to 100~mK. We find a broad peak in specific heat which is
 insensitive to magnetic field at a temperature of 1.5~K with a
 value of 2.65~J/mol~K. The magnetic susceptibility at 5~T continues to increase down to 2~K,
 the lowest temperature measured. The data suggest that the Sm spin system may be an ideal
 realization of the frustrated Heisenberg antiferromagnet on the square lattice.
\end{abstract}

%%%%%%%%%use  the \KEY command at the begin of keyword text%%%%%%%%%
%Please supply one or more relevant PACS-1996 classification codes
%(http://publish.aps.org/PACS/96pacs.htmland) and about 5 keywords
%of your own choice for indexing purposes.
%You can see a list of already used keywords for JMMM at
%http://authors.elsevier.com/JournalDetail.html?PubID=505704&Precis=KIND

\end{frontmatter}

\section{Introduction}\label{}

The ideally frustrated 2-D Heisenberg antiferromagnet with first
($J_1$) and second ($J_2$) nearest neighbor interactions on the
square lattice has been heavily studied
theoretically,\cite{ChandraPRB1988} but lacks few good examples in
nature. For small $J_2$/$J_1$ the system orders into a N\'{e}el
state, while for large $J_2$/$J_1$ one expects collinear order. At
$J_2$/$J_1 \approx$ 0.5 a spin liquid state whose properties are
not well known is expected. Experimentally, the best examples of
the spin 1/2 frustrated Heisenberg antiferromagnet on the square
lattice occur in the vanadates, such as
Li$_2$VO(Si,Ge)O$_4$,\cite{MelziPRL2000}
VOMoO$_4$,\cite{CarrettaPRB2002} and
Pb$_2$VO(PO$_4$)$_2$\cite{Kaul2004} where it is believed that
$J_2$/$J_1$ $>$ 1.

Here we report preliminary thermodynamic measurements on a single
crystal cuprate Sm(La,Sr)CuO$_4$. By alternately stacking SmO and
(La,Sr)CuO$_3$ layers this so called T$^{\star}$ structure of the
cuprates possesses 2-D Sm spin layers which are well isolated from
one another.\cite{TokuraPRB1989,FiskTstarcuprate1989}

\section{Results}

Figure 1 presents raw specific heat data from a quasiadiabatic
heat pulse method for Sm(La,Sr)CuO$_4$. At these temperatures the
phonon contribution which becomes dominant above $\sim$ 10~K is
negligible. There is a peak at $T_{max}$ = 1.5~K with a value of
$C(T_{max})$ = 2.65 J/mol K. The low temperature ($T <$ 0.2~K)
specific heat is the sum of a magnetic contribution and a nuclear
Schottky contribution. We subtract the nuclear Schottky
contribution that we model as the sum of a constant quadrupolar
term and a dipolar term subject to Zeeman splitting:
$C_{nuc}(T,H)$ = (0.0041 J K/mol + 1.3$~$10$^{-5} H^2$ J K/mol
T$^2$)/$T^2$. The resulting magnetic contribution to the specific
heat is shown in figure 2. Note that there remains a low
temperature upturn, which is suppressed with increasing magnetic
field, but no clear long range magnetic order is observed down to
100 mK.

 \begin{figure}[t]
\includegraphics[scale =0.9]{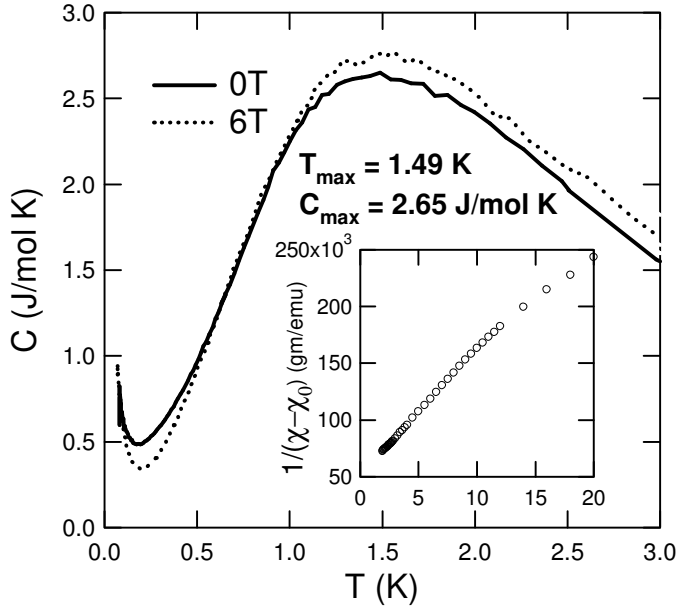}
\caption{Specific heat of Sm(La,Sr)CuO$_4$ in zero and applied
magnetic field in the ab-plane. (inset) Inverse susceptibility at
5~T with field in the ab-plane.}
    \label{fig-1}
  \end{figure}

The zero field cooled susceptibility shows a superconducting
transition at 15~K. By applying a field of 5 T in the ab-plane the
evidence for superconductivity is suppressed, and the
susceptibility continues to rise down to 2~K, the lowest
temperature measured. A Curie-Weiss plus constant fit to room
temperature allows us to extract a background paramagnetic
contribution $\chi_0$ = 2.4$~10^{-6}$ emu/gm. The remaining signal
at low temperature is attributed to the susceptibility of the Sm
spins and a Curie-Weiss fit below 10~K gives $\Theta_{CW}$ =
-4.5~K.

  \begin{figure}[t]
\includegraphics[scale =0.9]{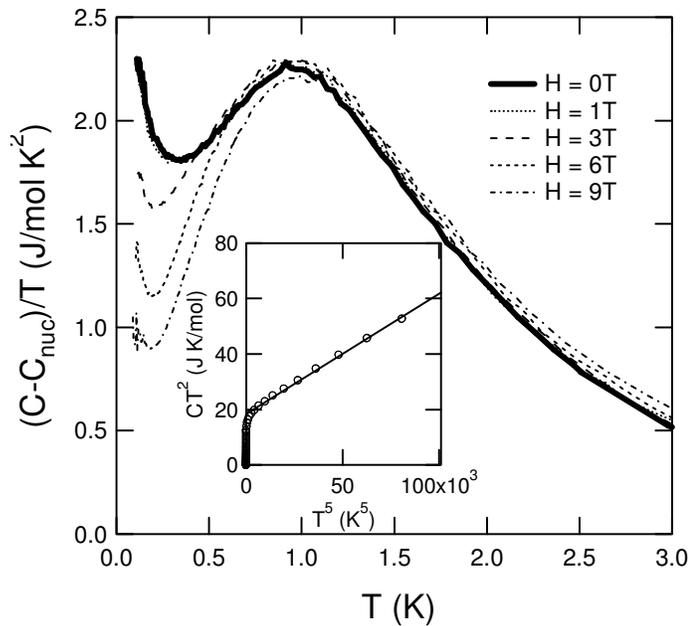}
\caption{Magnetic contribution to the specific heat of
Sm(La,Sr)CuO$_4$ after subtraction of the nuclear Schottky
contribution in zero and applied magnetic fields in the ab-plane.
(inset) Zero field specific heat data plotted to extract the high
temperature magnetic contribution. The solid line is a linear fit
from 4 to 10~K.}
    \label{fig-2}
  \end{figure}

\section{Discussion}

In the absence of frustration, the magnetic susceptibility should
show a peak at 0.935~$J$,\cite{KimPRL1998} where $J$ can be
determined by the low temperature Curie-Weiss fit. Therefore, we
have $T^{\chi}_{max}/\Theta_{CW} < 0.45$ which is strong evidence
for the presence of frustration. The small peak value of the
specific heat $C(T_{max})$ = 0.32 R, also suggests that
frustration is playing a key role in the Sm spin dynamics. Knowing
$C(T_{max})$ and $T_{max}$ we can use the work of Misguich, Bernu,
and Pierre to solve graphically for $J_1$ and
$J_2$.\cite{MisguichPRB2003} The two solutions are $J_2$/$J_1$ =
2.0~K/4.8~K = 0.42 and $J_2$/$J_1$ = 3~K/3~K = 1, which indeed
places this system very close to the spin liquid regime.
Meanwhile, the high temperature expansion of the $J_1$-$J_2$ model
predicts that the magnetic excitations should fall off as $C_{mag}
\approx 3J^{2}_{2D}R/8T^2$ where $J_{2D}$ = $(J^{2}_{1} +
J^{2}_{2})^{1/2}$ and $R$ = 8.314 J/mol K. By assuming that the
phonon contribution to the specific heat varies as $T^3$ up to
10~K, we can determine $J_{2D}$ by the $T$=0 linear extrapolation
from a plot of $CT^2$ versus $T^5$ as done in the inset of figure
2. We find $J_{2D} \approx$ 2.4~K. This value is roughly a factor
of 2 too small for either graphical solution found using the work
of reference \cite{MisguichPRB2003}. The graphical solution also
appears to over estimate $J_1$ and $J_2$ when considering that the
susceptibility also gives us $J_1$ + $J_2$ = $\Theta_{CW}$ =
4.5~K. These small discrepancies might be reconciled if there is
an additional mechanism, aside from a simple frustration model,
that reduces the peak height in the specific heat. The graphical
solutions could then lean towards smaller $J_1$ and $J_2$, with
$J_2$/$J_1 >$ 1. Whether or not additional longer range
interactions, such as the RKKY interaction which could be mediated
through the CuO$_2$ conduction layers, could achieve this remains
to be seen.

The low temperature upturn in $C/T$ in figure 2 may indicate the
onset of ordering either from 3-D coupling or an Ising like
transition expected in the limit that $J_2$/$J_1$ is large.

We should also caution that this system has the obvious added
complication of being embedded in a high temperature
superconductor, with $T_c(H=0)$ = 15 K as determined from a
susceptibility measurement. While this fact could be used to
extract the magnetic spectrum via transport
measurements,\cite{BulaevskiiPRL2005} it may also have a
significant effect on the Sm-Sm exchange interaction as observed
previously in several other cuprates containing rare-earth
elements within the charge reservoir layers.\cite{Allenspach2000}

\section{Acknowledgements}
We are very grateful to C. Batista for fruitful discussions. Work
at Los Alamos was performed under the auspices of the US DOE.

%%%main text

%%%%%%%%%%%%%%%%%%%% Kondo in title, abstract and/or keywords %%%%%%%%%%%%%%%%

%%%%%%%%%%%%%%%%%%%%%%%%%%%%%%%%%%%%%%%%%%%%%%%%%%%%%%%%%%%%%%%%%%%%%%%%%%%%%%

\end{document}